\documentclass[twocolumn,showpacs]{revtex4}

\usepackage{bm}% bold math
\usepackage{amssymb}
\usepackage{amsmath}
\usepackage{graphics}
\usepackage[sort&compress]{natbib}
\usepackage{epsfig}
\usepackage{color}

\usepackage[ps2pdf,colorlinks=true, pagebackref=false, bookmarks=true,bookmarksopen=true,bookmarksnumbered=true]{hyperref}

\newcommand{\Cu}{\ensuremath{\mathrm{Cu}}}
\newcommand{\Ni}{\ensuremath{\mathrm{Ni}}}
\newcommand{\Nb}{\ensuremath{\mathrm{Nb}}}
\newcommand{\Al}{\ensuremath{\mathrm{Al}}}

\renewcommand{\O}{\ensuremath{\mathrm{O}}}

\renewcommand{\section}[1]{}
\renewcommand{\subsection}[1]{}

\begin{document}

%\preprint{}

\title{Magnetic anisotropy in ferromagnetic Josephson junctions%
}

\author{M. Weides}
\affiliation{%
  Center of Nanoelectronic Systems for Information Technology and Institute of Solid State Research, Research Centre J\"ulich, D-52425 J\"ulich, Germany%
}

\date{\today}% It is always \today, today,
             % but any date may be explicitly specified

\begin{abstract}
Magnetotransport measurements were done on $\Nb/\Al_2\O_3/\Cu/\Ni/\Nb$ superconductor-insulator-ferromagnet-superconductor Josephson tunnel junctions. Depending on ferromagnetic $\Ni$ interlayer thickness and geometry the standard ($1d$) magnetic field dependence of critical current deviates from the text-book model for Josephson junctions. The results are qualitatively explained by a short Josephson junction model based on anisotropy and $2d$ remanent magnetization.
\end{abstract}
\pacs{%
  74.25.Fy
  74.45.+c %Proximity effects; Andreev effect; SN and SNS junctions
  74.50.+r, %Proximity effects, weak links, tunneling phenomena,
              %and Josephson effect
  74.70.cn
% 74.78.Db %Low-Tc films
% 75.45.+j, %Macroscopic quantum phenomena in magnetic systems
% 85.25.Cp %Josephson devices
}

\keywords{%
  Josephson junctions, $\pi$-junction, Superconductor ferromagnet superconductor junctions
}%Use showkeys class option if keyword display desired

\maketitle

%\section{Introduction}
Superconductivity (S) and ferromagnetism (F) in thin layered films have now been studied during some decades \cite{buzdin05RMP}. In SF bilayers the superconductivity may be
non-uniform \cite{Demler97}, i.e. the Cooper pair wave function extends to the ferromagnet with an oscillatory behavior. In Josephson junctions (JJs) based on $s$-wave superconductors the phase coupling between the superconducting electrodes can be shifted by $\pi$ when using a ferromagnetic barrier with an appropriate chosen thickness $d_F$, i.e. SFS or SIFS-type junctions (I: insulating tunnel barrier).
Only in recent years the experimental realization of $\pi$ JJs was successful. In particular, the $\pi$ coupling was demonstrated by varying the temperature \cite{Ryazanov01piSFS_PRL,Sellier03TinducedSFS,WeidesHighQualityJJ}, the thickness of
the F-layer \cite{Blum:2002:IcOscillations,Kontos02Negativecoupling,WeidesHighQualityJJ,OboznovRyazanov06IcdF} or measuring the
current-phase relation of JJs incorporated into a superconducting
loop \cite{BauerHall,Ryazanov2001piSquid,GuichardApril03DCSquid}. The coupling can also change within a single JJ by a step-like F-layer, i.e. one half is a $0$ JJ and the other half is a $\pi$ JJ \cite{WeidesFractVortex,WeidesSteppedJJ}.\\For useful classical or quantum circuits based on SFS/SIFS JJs a large critical current density $j_c$ (small Josephson penetration depth $\lambda_J$) and a high $I_cR$ product are needed \cite{ustinovRapidsingle-fluxquantumlogicusingpi-shifters,KatoPRB07}. Up to now the limiting factor is the low $j_c$ due to strong Cooper pair breaking inside F-layer. Alloys of magnetic and non-magnetic atoms such as $\Ni\Cu$ face problems of clustering \cite{houghton70} and strong magnetic scattering \cite{OboznovRyazanov06IcdF,WeidesHighQualityJJ}. Promising experiments using strong ferromagnet transition metals \cite{ShelukhinSFS06,RobinsonBlamire,Blum:2002:IcOscillations,Bell2005,RobinsonPRB07} were published.\\ Shape anisotropy of magnetic interlayer may provoke a not flux-closed domain structure and consequently a shift of critical current diffraction pattern $I_c(H)$ \cite{Bell2005}. In experiments \cite{Ryazanov01piSFS_PRL,Kontos02Negativecoupling,WeidesHighQualityJJ,Bell2005,RobinsonPRB07} the SFS/SIFS JJs had nearly mirror-symmetrical $I_c(H)$, i.e. the effective shift along $H$-axis is small, usually less than one flux quantum $\Phi_0$. This is explained by a multi-domain state of F-layer with a very small net magnetization. However, up to now the $2d$ nature of thin-film magnetism was disregarded.\\In this Letter the $I_c(H)$ dependence for remanent $2d$ magnetization of F-layer is systematically studied. First, the maximal flux from F-layer is estimated. Second, the $I_c(H)$ pattern considering $2d$ in-plane magnetization is calculated for different aspect ratios. Third, the $I_c(H)$ pattern is measured along both magnetic axes for various junction geometries and $d_F$.

\begin{figure}[t]
\includegraphics[width=8.6cm]{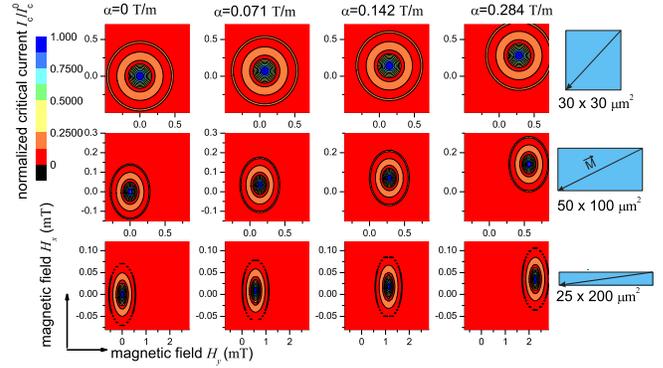}
  \caption{(Color online) Calculated surface plot of $I_c(H_x,H_y)$ for different geometries and remanent magnetizations $\alpha$. The magnetization vector $\protect\vec{M}$ points from top right to bottom left corner (arrow), $I_c^0$ is shifted in opposite direction.}
  \label{IcHSimu}
\end{figure}

\begin{figure}[t]
\includegraphics[width=8.6cm]{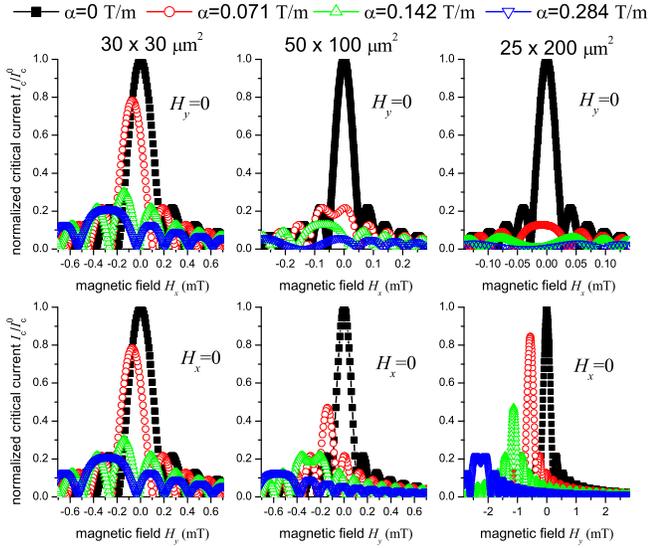}
  \caption{(Color online) Calculated $I_c$ as function of $1d$ magnetic field, $I_c(H_x,0)$ (top) and $I_c(0,H_y)$ (bottom) for the three geometries and various $\alpha$. A substantial deviation from ideal ($\alpha=0\:\rm{T/m}$) pattern appears for already small magnetization vector $\vec{M}=\alpha(L_x,L_y)$.}
  \label{IcHxySimu}
\end{figure}

\begin{figure}[t]
\includegraphics[width=8.6cm]{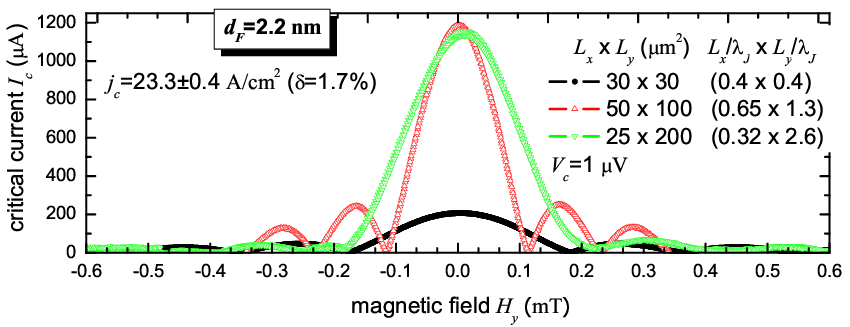}
\includegraphics[width=8.6cm]{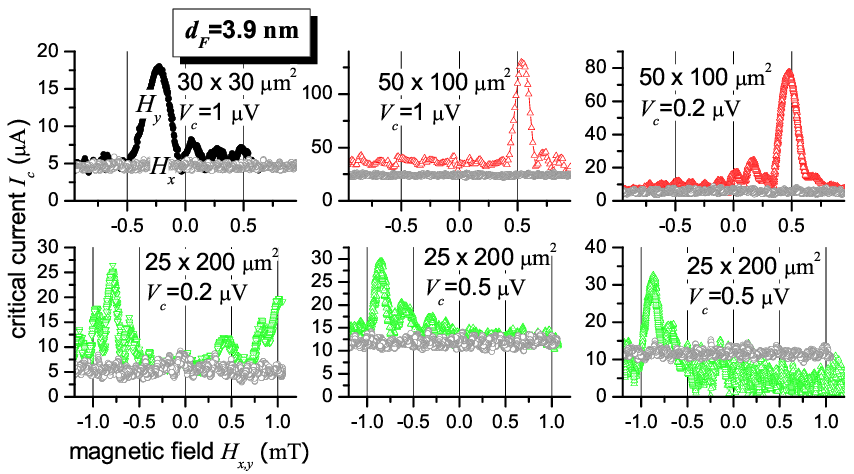}
  \caption{(Color online) Measured $I_c(H_y)$ pattern of SIFS JJs for thin (top) and thick (bottom) $\Ni$ layer and different geometries. $I_c(H_x)$ pattern of thick $\Ni$ layer is plotted in gray. Onset of magnetic anisotropy effects in $\Ni$ layer is between $2.2$ and $3.9\:\rm{nm}$.}
  \label{IcH}
\end{figure}
%\section{Theory}
The maximal shift of $I_c(H_y)$ is estimated for a strong magnet, i.e. $\Ni$, being magnetized fully in-plane and along $y$-axis. The atomic magnetic momentum is $0.6\:\mu_{\rm{B}}$ \cite{AhernNiCu}, the specific density $\rho$ is $8.9\:\rm{g/cm^3}$ (bulk) and magnetization $\mu_0M=0.64\:\rm{T}$. A cross-section of length $L_x=100\:\rm{\mu m}$ and F-layer thickness $d_F=3\:\rm{nm}$ encloses a magnetic flux $\Phi_M=d_F L_x\mu_0M$. The total magnetic flux $\Phi$ through the JJ is the applied field flux $\Phi_H=(2\lambda_L+d_F) L_x H_y$ (London penetration depth $\lambda_L=90\:\rm{nm}$) plus $\Phi_M$, i.e. $\Phi=\Phi_H\pm\Phi_M=8.85\Phi_0H_y/\rm{mT}\pm 92\Phi_0$.
The $I_c(H_y)$ pattern is shifted by $92$ periods from the center. This simple calculation neglects dead magnetic layer \cite{LiebermannDeadLayer70}, as found in SFS/SIFS JJs \cite{WeidesHighQualityJJ,RobinsonPRB07,OboznovRyazanov06IcdF}, and demagnetizing by domains. \emph{Real} $\Ni$ films tend to form complex magnetization profiles (in/out-of-plane) and domain structures as function of $d_F$ \cite{GubbiottiPRB02}. Integral magnetization measurements in SF layers show a complex behavior as a function of temperature, applied field, and sample history \cite{MontonPRB08,JoshiNbNi07}, e.g. SF structures spontaneously alter their stray field by changing magnetic domain distribution \cite{DubonosPRB02}. The local magnetization depends on stray fields from neighbor domains, flux focusing from S-electrodes \cite{Bell2005,Wu_SFlayerPRB07} or on bias induced spin accumulation at F/S interface \cite{JedemaPRB99}. It is generally agreed that the average magnetization in SFS/SIFS JJs is much smaller than the maximum magnetization estimated above. But as shown below even a remanent $2d$ magnetization of $1\%$ of the maximal value may notable change the $I_c(H)$ pattern.

A qualitative model for $I_c(H)$ in presence of a uniform, fixed $2d$ magnetization $\vec{M}$ is derived. The short JJ model \[I_c(H)=I_c^0\left|\frac{\sin(\pi\frac{\Phi}{\Phi_0})}{\pi\frac{\Phi}{\Phi_0}}\right|\quad \mathrm{and}\quad\Phi=\Phi_H \pm \Phi_M\] is modified by $2d$ distributions of applied field flux $\Phi_H$ and magnetization $\Phi_M$ with \[\Phi=\left|\left[(2\lambda_L+d_F)\left(\begin{array}{cc}0&H_x\\H_y&0\end{array}\right)\pm\left(\begin{array}{cc}\alpha&0\\ 0&\alpha\end{array}\right)\right]\times\left(\begin{array}{c}L_x\\L_y\end{array}\right)\right|.\] $\vec{M}=\alpha(L_x,L_y)$ is assumed to be orientated in-plane (Meissner-screening of S-electrodes) along the diagonal (i.e. longest) axis of sample (rough approximation for magnetic shape anisotropy). The magnetization $|\vec{M}|$ is several orders of magnitude smaller than the upper limit given by a fully saturated magnetic layer. Note that the easy axis of ferromagnetic film can be determined by the magnetic field during deposition, too. The model for $\Phi_M$ is just exemplary for the effect of $2d$ in-plane magnetization in $I_c(H)$.\\ In Fig. \ref{IcHSimu} the surface plot of $I_c(H_x,H_y)$ is depicted for various $\alpha$ and geometries. The position of $I_c^0$ is shifted from the center ($H_x=H_y=0\:\rm{mT}$) in opposite direction of $\vec{M}$. Fig. \ref{IcHxySimu} depicts $I_c(H_x,0)$ and $I_c(0,H_y)$ pattern. These graphs resemble standard $1d$ $I_c(H)$ ($H=H_x,H_y$) measurements. For some $|\vec{M}|\neq0\:\rm{T}$ a single-peaked $I_c(H_x)$, $I_c(H_y)$ pattern is calculated, which -on first glance- resembles a shifted $|\sin(H)/H|$ Fraunhofer pattern. However, the maximum $I_c$ is smaller than the real $I_c^0$ and the height of side-maxima do not obey the expected value. For example the $I_c(H_x,0)$ pattern of $30\times30\:\rm{\mu m^2}$ sample and $\alpha=0.071\:\rm{T/m}$ is shifted by less than $\Phi_0$ and its maximum $I_c$ is already reduced to $\sim0.8I_c^0$. This simple $I_c(H_x,H_y)$ model may qualitatively explain some experimental observations (Fig. \ref{IcH}) on SIFS-type JJs with $\Ni$ interlayer.

For experiment JJs with similar areas, i.e. $30\times30, 50\times100 \;\rm{and}\; 25\times200\:\rm{\mu m^2}$, were fabricated.
The deposition and structuring of JJs is described in Ref. \onlinecite{WeidesFabricationJJPhysicaC}. The SIFS multilayer was magnetron sputter deposited with $\Ni$ thickness $d_F$ ranging from $1\rm{-}6\:\rm{nm}$. The tunnel barrier was formed for $30\:\rm{min}$ at a partial oxygen pressure of $0.1\:\rm{mbar}$. After oxidation a $2\:\rm{nm}$ $\Cu$ film was inserted. All JJs were deposited in a single run by shifting the substrate and target to obtain a wedge-shaped $\Ni$-layer. Normal state and subgap resistance indicate a small junction to junction variation. The IV and $I_c(H_x)$, $I_c(H_y)$ characteristics were measured at $4.2\:\rm{K}$ for two sets of samples ($d_F=2.2\,,\,3.9\:\rm{nm}$). Cooldown was done in zero field and thermal cycling up to $\approx15$ and $300\:\rm{K}$ to check reproducibility. Transport measurements were made in a liquid He dip probe using low-noise home made electronics and room-temperature voltage amplifier. The magnetic fields $(H_x,0)$, $(0,H_y)$ were applied in-plane of the sample and parallel the junctions axis (Fig. \ref{IcHSimu}). The voltage criteria $V_c$ for $I_c(H)$ determination was $0.2\rm{-}1\:\rm{\mu V}$. A lower subgap resistance for $d_F=3.9\:\rm{nm}$ sample leads to larger offset currents. Positive and negative current branch of IVC had similar magnetic field dependence $+I_c(H_y)\simeq|-I_c(H_y)|$. Magnetic field was swept between $\pm1.5\:\rm{mT}$ for all samples. All junctions had their lateral sizes comparable or smaller than the Josephson penetration length $\lambda_J$, except the longest sample ($d_F=2.2\:\rm{nm}$, $25\times200\:\rm{\mu m^2}$), whose $L_y$ is not strictly inside the short JJ limit.\\The $d_F=2.2\:\rm{nm}$ samples showed very regular $I_c(H_y)$ pattern. All maximum $I_c$'s were nearly centered and the spread of $j_c$ was $\sim1.7\%$, as determined from the maximum $I_c$'s. The $I_c(H_x)$ pattern is symmetric, too (not shown). The oscillation period of $I_c(H_y)$ were determined by magnetic cross-section $\sim 1/L_x$, and nearly independent of aspect ratio. No indication for a distorted supercurrent transport due to alloying at the $\Nb/\Ni$ interface \cite{ChenNbNi} can be found. Effects due to magnetic anisotropy were not detectable, either because the samples were still inside the dead magnetic layer, or the anisotropy was absent or totally out-of-plane. \\The $d_F=3.9\:\rm{nm}$ samples had completely different $I_c(H_y)$ pattern showing in-plane magnetic anisotropy with some characteristic features. All maximum $I_c$'s were shifted from the center, and the amplitude of shift increased with $L_y$, i.e. $\approx0.24\:\rm{mT}$ for $30\times30\:\rm{\mu m^2}$, $\approx0.5\:\rm{mT}$ for $50\times100\:\rm{\mu m^2}$, and $\approx0.8\:\rm{mT}$ for $25\times200\:\rm{\mu m^2}$ samples. The direction of  shift varies between samples even if they were cooled and measured at the same time (random polarity of magnetic configuration). The position of main peak of $I_c(H_y)$ was reproducible after thermal cycling to $300\;\rm{K}$. The width of main maxima (measured at offset line) was not strictly  $\sim1/L_x$, and varies from sample to sample. The pattern were asymmetric, i.e. the height of same-order side maxima differed, probably due to non-uniform flux guidance in F-layer and re-orientation of domains.\\By rotating the magnetic field by $90^\circ$, i.e. measuring in $I_c(H_x)$ mode, low $I_c$'s, being nearly independent of $H_x$, were detected. Even the squared shaped $30\times30\:\rm{\mu m^2}$ JJ had an almost flat $I_c(H_x)$ pattern. This indicates some magnetic crystallographic anisotropy along $y$-axis, probably caused by magnetron sputter deposition. Small deviations of $I_c(0,0)$ for $I_c(H_x,0)$ and $I_c(0,H_y)$ measurements can be related to variations of magnetic configuration by the unshielded sample handling at $300\:\rm{K}$. A considerable spread of maximum $I_c$ (Fig. \ref{IcH}) can be already seen for JJs with same geometry, which is even increased by considering the maximum $j_c$ for different geometries. Simulations (Fig. \ref{IcHxySimu}) show that already a moderate magnetization $\vec{M}$ ($\alpha<0.1\:\rm{T/m}$) yields very different maximum $I_c$'s of $I_c(H_x)$ and $I_c(H_y)$. A sample to sample variation of direction and amplitude of $\vec{M}$ explains the data spread of Fig. \ref{IcH}. Obviously, this leads to very large variations in the $I_c(d_F)$ dependence.\\In literature, the $I_c(H)$ pattern of SFS/SIFS JJs with \emph{comparable} strong magnets were either shown for samples with thin $d_F$ \cite{Bell2005}, or had deviations from the ideal $|\sin(H)/H|$ form. For example, the maximum $I_c$ in Ref. \onlinecite{RobinsonPRB07}, Fig. 5 (inset) is too small compared to the first side-maxima. These samples were small in area ($\leqslant1\:\rm{\mu m^2}$), and the F-layer could have been in single domain state. As $I_c$ varied smoothly with $H$, either $\vec{M}$ rotated softly, or a multi-domain structure with averaged $2d$ magnetization existed. For both cases the $2d$ nature of remanent magnetization may have suppressed the maximum $I_c$ -determined from $I_c(H)$ pattern- below the real $I_c^0$.

In summary, the $I_c(H)$ pattern along both field axis of SIFS JJs with $\Ni$ interlayer were measured. Assuming magnetic anisotropy the characteristic features, i.e. shift or absence of central peak, can be qualitatively reproduced by simulations. As conclusion, the $1d$ $I_c(H)$ pattern in presence of magnetic anisotropy can not yield the \emph{real} $I_c^0$. Future experiments on SFS/SIFS JJs should be done by $2$-axis $I_c(H_x,H_y)$ scan of JJs with well-controlled magnetic interlayer domain configuration. Superconducting spintronic devices such as FSF spin valves and $\pi$, $0$--$\pi$ JJs are promising candidates for future classical and quantum computers. As shown by this Letter control of magnetic anisotropy and magnetic domain configuration is essential for phase-switchable FSF/SFIFS or SFS/SIFS devices.\par
M.W. thanks D. Sprungmann for proof-reading, A. Bannykh for help with fabrication and DFG for support (project WE 4359/1-1).

%\bibliographystyle{apsprl}
%\bibliography{G:/Juelich/eigene_Paper/BIB}
%\bibliography{C:/Juelich/eigene_Paper/BIB}
%\bibliography{X:/user2/weides/eigene_Paper/BIB}

%\end{document}

\end{document}